# Studies of magnetogyration in cubic $Bi_{12}GeO_{20}$ crystals using small-angular magnetooptic polarimetry

Krupych O., Adamenko D., Say A., Klymiv I., Vlokh R. and Vlokh O.

Institute of Physical Optics, 23 Dragomanov St., 79005 Lviv, Ukraine,
e-mail: vlokh@ifo.lviv.ua

## Abstract

We present the results of studies for magnetogyration (MG) effect in non–centro-symmetric, non-polar $Bi_{12}GeO_{20}$ crystals performed with the small–angular polarimetric mapping technique. The magnitude of MG rotation obtained with the small–angular mapping is close to the corresponding values obtained recently on the basis of single–ray polarimetry. Our results demonstrate that multiple reflections of light play a key role in the studies of MG effect in $Bi_{12}GeO_{20}$ crystals. They lead to the error, which is at least two times as large as the expected value of the MG rotation.

## Introduction

In the recent works [1,2] we have demonstrated a significance of small-angular imaging polarimetric techniques for studies of magnetooptic phenomena such as Faraday rotation (FR) and magnetogyration (MG) (see [3]). It seems for us that the MG effect should not be forbidden in the media, which stay in non-equilibrium conditions [4]. In such a case the existence of this effect would not contradict the well-known Onsager principle.

In order to extract MG from a total magnetooptic rotation (MOR), a specimen is usually examined with two light beams traversing in the opposite directions. Then we have $MOR^+ = FR + MG$ for one direction and $MOR^- = FR - MG$ for its reverse. The difference $MOR^+ - MOR^-$ should give us the MG effect.

We have shown on the examples of CdS and $(Ga_{0.3}In_{0.7})_2Se_3$ crystals that the results [5,6] reported previously on the MG effect, which have been obtained with a single-ray polarimetry, are not corroborated. This fact is caused by the existence of some unsuspected experimental errors, along with an extreme smallness of the MG effect. Eventually, the differences between the MOR values measured under the wave vector reversal (which could be, in principle, interpreted as the MG) are close to the experimental error.

Both CdS and $(Ga_xIn_{1-x})_2Se_3$ crystals are polar, wide-band semiconductor crystals with significant absorption of light in the visible spectral range. CdS and $(Ga_xIn_{1-x})_2Se_3$ belong to the point groups of symmetry 6mm and 6, respectively. It means that both materials are anisotropic and optically uniaxial. Therefore, the MOR could be observed directly only if the light beam and the magnetic field are oriented along the optic axis. Otherwise, linear birefringence effect would distort the experimental results. That is the reason why misalignments of crystalline samples in the case of CdS and $(Ga_xIn_{1-x})_2Se_3$ could introduce additional errors into the final results for the MG.

Unlike CdS and $(Ga_xIn_{1-x})_2Se_3$ crystals, $Bi_{12}GeO_{20}$ (abbreviated hereafter as BGO) is cubic (the point symmetry group 23), non-centrosymmetric and non-polar absorbing crystal. This implies that the linear birefringence is zero and it could not affect the results for the MOR. Moreover, the BGO crystals are well-known photorefractive materials (see, e.g., [7,8]). Due to appearance of photovoltaic current, the crystals can very likely stay in a non-equilibrium state under optical irradiation. In the previous work [9], a single-ray polarimetry has been used for studying MG in the BGO crystal. The objective of the present article is to restudy MG effect in $Bi_{12}GeO_{20}$ crystals by means of small-angular magnetooptic polarimetry.

**Experimental**

Small-angular imaging polarimetric technique was earlier described in detail in [1,2]. We used in our studies the imaging polarimetric setup presented in Fig. 1. The difference from a conventional imaging polarimetry (see, e.g., [10]) consists in the use of conical probing beam, instead of a parallel one. The dimensions of the light channel in the magnetic core 7 limited the angular divergence of the conical probing beam. We used He-Ne laser ($l = 632.8 nm$) as a source of optical radiation. Sample 8 was positioned at the beam waist. Objective lens 10 pictured the cross-section of the light beam passed analyzer 9 onto the sensor of CCD camera 11. The image obtained by the camera corresponded to the angular aperture of about $3.49 \times 10^{-2}$ rad (or ~ 2 deg). That is why this technique was called as "small-angle polarimetric mapping".

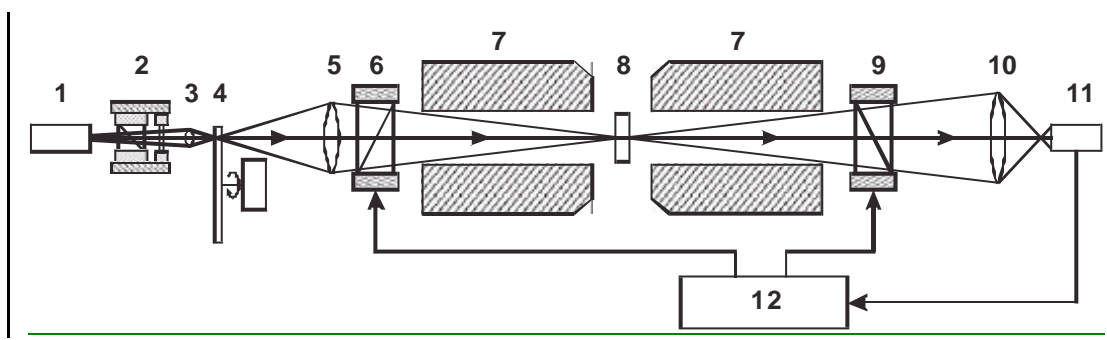

**Fig. 1.** Optical scheme for small-angle magnetooptic polarimetric mapping: 1 – laser; 2 – circular polarizer (linear polarizer and a quarter-wave plate); 3 – short-focus lens; 4 – coherence scrambler; 5 – long-focus lens; 6 – linear polarizer (Glan prism) with motorized rotary stage; 7 – magnetic core; 8 – sample; 9 – analyzer (Glan prism) with motorized rotary stage; 10 – objective lens; 11 – CCD camera; 12 – computer.

The crystalline sample was placed between the poles of electromagnet. The sample was a plane-parallel plate of (111) crystallographic cut, with the thickness of $d = 6.17$ mm. The distance between the electromagnet cores (54 mm) was large when compare with the sample thickness, thus allowing us to reduce inhomogeneities of the magnetic field (and the appearance of transverse component of that field) through the sample thickness to a negligibly small value. The sample was oriented by aligning the beam reflected from its face with the incident light beam.

The small-angular maps of the polarization azimuth were obtained in the absence of magnetic field and for the case of the magnetic field $H_z = 4.63$ kOe, for the two orientations of sample (a 'direct' one and that corresponding to the sample rotated by 180°). Basing on these maps, we calculated the MORs for the both wave vector directions +$k$ and –$k$.

### Results and discussion

The small-angular maps of polarization azimuth for the BGO crystal are shown in Fig. 2. The MOR is a difference between the azimuths detected in the presence ($\boldsymbol{b}_H$) and absence ($\boldsymbol{b}_0$) of the magnetic field and so it could be calculated as

$$\Delta \boldsymbol{b} = \boldsymbol{b}_H - \boldsymbol{b}_0 . \qquad (1)$$

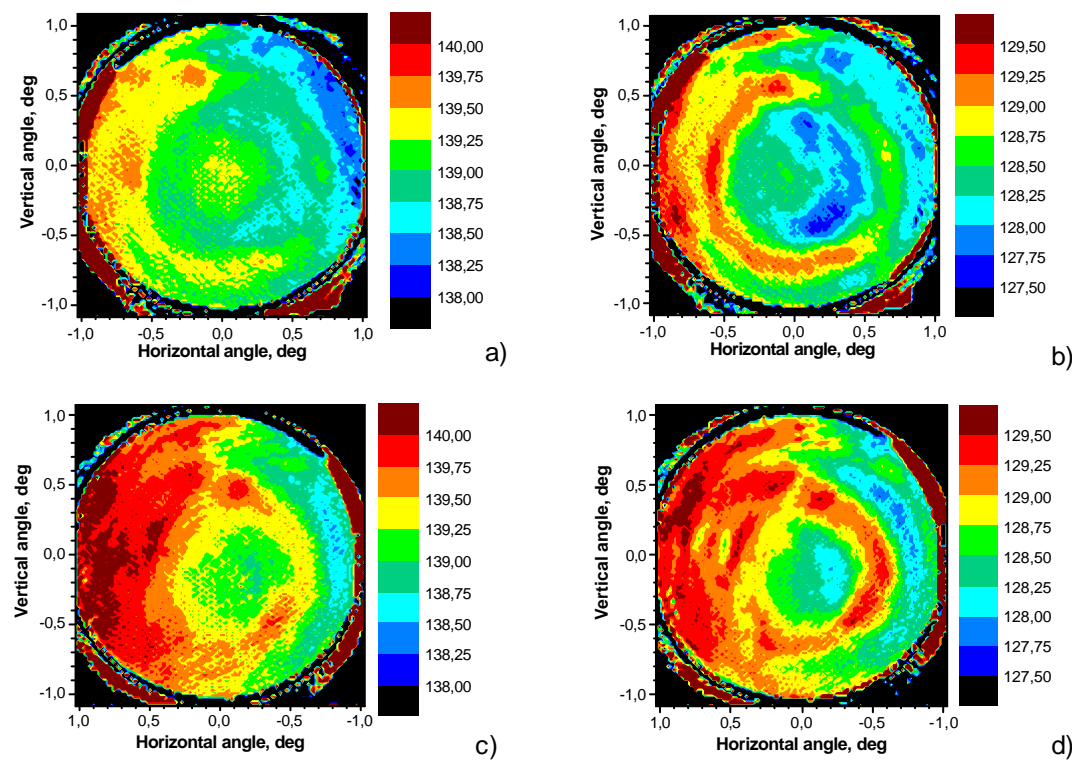

**Fig. 2.** Maps of polarization azimuth (in angular degrees) for the BGO crystals: (a) +$k$, $H = 0$, (b) +$k$, $H = 4.63$ kOe, (c) –$k$, $H = 0$ and (d) –$k$, $H = 4.63$ kOe.

The maps of the mentioned difference (i.e., the maps of the MOR) are represented in Fig. 3. We have marked the circular regions in the centre of the MOR maps with the angular dimension of $0.23° = 4$ mrad, which correspond to a divergent laser beam used in the single-ray polarimetry. Next, we have calculated the mean values of the MOR for the chosen regions and the corresponding experimental accuracies: $\Delta b^+ = 10.943° \pm 0.228°$ and $\Delta b^- = 10.788° \pm 0.197°$. Using these values and the formula

$$a_{33} = \frac{l n_o r_3}{p H_3}, \qquad (2)$$

we have calculated the coefficient of the Faraday effect: $a_{11} = (340.3 \pm 9.4) \times 10^{-11} Oe^{-1}$. The latter is close to the value obtained earlier with the single-ray polarimetric technique ($a_{11} = 321.3 \times 10^{-11} Oe^{-1}$ [9]).

Non-reciprocal MOR (abbreviated as NRMOR) is a difference of MORs ?β for the opposite directions of the wave vector (+k and –k):

$$d(\Delta b) = \Delta b^+ - \Delta b^- . \qquad (3)$$

We have calculated the NRMOR map (see Fig. 4) as a difference of the maps depicted in Fig. 3. For the central region, we have calculated the mean value of the NRMOR and the corresponding experimental accuracy: $d(\Delta b) = 0.155° \pm 0.301°$. It is known that the NRMOR should be interpreted as a manifestation of MG effect.

MG coefficient calculated from the above NRMOR value is equal to $d_{321} = (2.4 \pm 4.7) \times 10^{-11} Oe^{-1}$. It is worthwhile to notice that the MG coefficient obtained by using single-ray polarimetric technique is $d_{321} = 3.5 \times 10^{-11} Oe^{-1}$ [9]. This value lies in the confidence interval of the coefficient obtained in the presented work. However, our present results show that the NRMOR magnitude and, consequently, the MG coefficient are nearly twice as less as the experimental error. As seen from the data of Fig. 2 to Fig. 4, the rings of optical rotation value are observed on the maps. Obviously, these rings are caused by multiple reflections of light from the interfaces of sample (see, e.g., [11,12]). Thus, the principal source of the error in our case is just the multiple reflections. The effect can be eliminated by introducing the sample into index-matching liquid or deposition of antireflection coating.

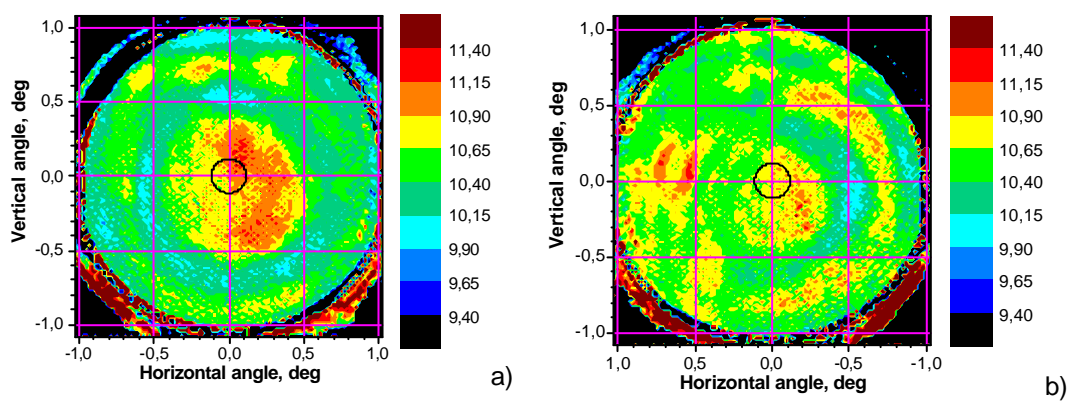

**Fig. 3**. Maps of MOR (in angular degrees) for the BGO crystals: (a) +k and (b) −k.

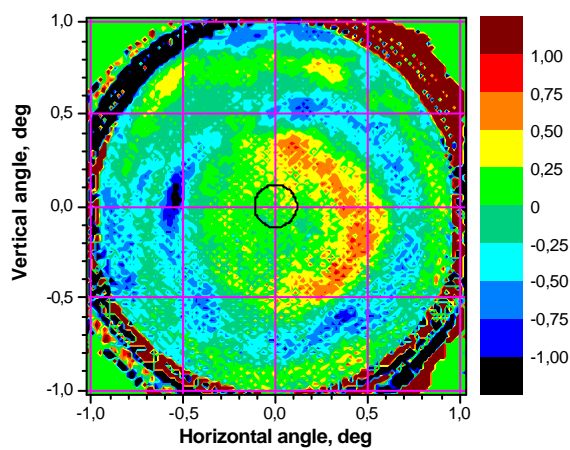

**Fig. 4**. Map of the NRMOR (in angular degrees) for the BGO crystal.

## Conclusion

The value of the NRMOR obtained in this work on the basis of small-angular polarimetric mapping is close to that obtained recently with the single-ray polarimetry. At the same time, the results presented above demonstrate that the studies of MG effect in the BGO crystals face a principal obstacle that represent the multiple reflections of light. The latter phenomenon could lead to the errors, which are at least two times larger than the expected value of MG rotation. The results of studies for the MG effect in BGO based on using index-matching liquid or antireflection coating will be presented in the forthcoming article.